\documentclass[12pt]{iopart}

\usepackage{iopams}
\usepackage{bbold}
\usepackage{graphicx}
\usepackage[breaklinks=true,colorlinks=true,linkcolor=blue,urlcolor=blue,citecolor=blue]{hyperref}

\usepackage{dsfont}
\expandafter\let\csname equation*\endcsname\relax
\expandafter\let\csname endequation*\endcsname\relax

\usepackage[utf8]{inputenc}
\usepackage[english]{babel}
\usepackage{braket}
\usepackage{amsmath}
\usepackage{graphicx}
\usepackage{multirow}
\usepackage{ulem} %added to be able to cross text out
\usepackage{units} % for using nicefrac in texts
\usepackage{version} % for comment
\usepackage{color}
\usepackage{colortbl}
\usepackage{xcolor}  % for writting parts in color, e.g. in red
\usepackage{amsmath}

\begin{document}

%\preprint{}


%\title{Local relaxation and velocity of correlations in a coherently split trapped one-dimensional Bose gas} 

\title{A marginally stable optical resonator for enhanced atom interferometry}

\author{I. Riou$^{1,5}$\footnotemark[1], N. Mielec$^{2,5}$\footnotemark[1],G. Lefèvre$^{1,5}$, M. Prevedelli$^{3}$, A. Landragin$^{2,5}$, P. Bouyer$^{1,5}$, A. Bertoldi$^{1,5}$, R. Geiger$^{2,5}$, and B. Canuel$^{1,4,5}$}
\footnotetext[1]{Both authors contributed equally to this work.}
\eads{\mailto{remi.geiger@obspm.fr},\mailto{benjamin.canuel@institutoptique.fr}}
\address{$^1$LP2N, Laboratoire Photonique, Numérique et Nanosciences, Université Bordeaux--IOGS--CNRS:UMR 5298, rue F. Mitterrand, F--33400 Talence, France.}
\address{$^2$LNE--SYRTE, Observatoire de Paris, PSL Research University, CNRS, Sorbonne Universités, UPMC Univ. Paris 06, 61 avenue de l’Observatoire, F--75014 Paris, France.}
\address{$^3$Dipartimento di Fisica e Astronomia, Universit{\`a} di Bologna, Via Berti-Pichat 6/2, I--40126 Bologna, Italy.}
\address{$^4$LSBB, Laboratoire Souterrain à Bas Bruit
UNS, UAPV, CNRS:UMS 3538, AMU, La Grande Combe, F--84400 Rustrel, France.}
\address{$^5$MIGA Consortium}

%\thanks{}

\date{\today}

\begin{abstract}
We propose a marginally stable optical resonator suitable for atom interferometry. The resonator geometry is based on two flat mirrors at the focal planes of a lens that produces the large beam waist required to coherently manipulate cold atomic ensembles. 
Optical gains of about 100 are achievable using optics with part-per-thousand losses. The resulting power build-up will allow for enhanced coherent manipulation of the atomic wavepackets such as large separation beamsplitters.
We study the effect of longitudinal misalignments and assess the robustness of the resonator in terms of intensity and phase profiles of the intra-cavity field. 
We also study how to implement atom interferometry based on Large Momentum Transfer Bragg diffraction in such a cavity.
\end{abstract}

%\pacs{XXXX}
%\noindent{\it Keywords}: in preparation, IOP journals
%\submitto{\NJP}
%\tableofcontents
%\maketitle

\section{Introduction}

Optical cavities have been extensively adopted in cold neutral atom physics to mediate the interaction between radiation and matter. For instance, quantum electrodynamics uses cavity to study the coupling of atoms with the light in an optical resonator  \cite{Walther2006,Haroche2013}, and cavity tailored spontaneous emission probability of atomic ensembles is at the basis of self-organization and superradiance phenomena \cite{Ritsch2013,Baumann2010}. Optical cavities have been used to produce Bose-Einstein condensates in the strong-coupling regime \cite{Colombe2007}, and to entangle particles of an atomic ensemble in order to beat the so-called standard quantum limit \cite{Cox2016,Hosten2016}. Recently, a matter-wave interferometer has also been realized using  an optical cavity, in a configuration where atoms are trapped in a small cavity mode volume \cite{Hamilton2015}.
%($600\ \mu$m)  

In atom interferometry, the optical cavity provides power enhancement and spatial filtering for the beamsplitters, with potential advantages in terms of both sensitivity and accuracy. The power enhancement can be exploited to increase the scale factor of atomic inertial sensors by using coherent multiphoton processes \cite{Muller2008b,Leveque2009,Chiow2011,McDonald2013} with large interrogation laser intensities. Such Large Momentum Transfer (LMT) techniques impose  constraints on the laser power, because of their velocity selectivity with respect to the Doppler effect experienced by the atoms at finite temperature. For example, a LMT beam splitter of order $n$ requires a laser intensity scaling as $n^4$ at fixed spontaneous emission rate \cite{Muller2008a}. 
The spatial filtering on the cavity modes can in addition reduce wavefront distortion \cite{Louchet-Chauvet2011}, an effect that can eventually limit the accuracy of atom interferometers, especially when using LMTs.

%Moreover, an interferometric control of the beam splitter wavefront could reduce the systematic effects introduced by beam distortions \cite{Louchet-Chauvet2011}; more precisely, using an optical cavity to filter the laser mode moves the problem to the quality of the mirrors implementing the resonator. 
%\textcolor{teal}{Is this argument relevant? ten lines later we say that misalignment in our resonator introduces phase distortion of the mode}

%\textcolor{red}{State-of-the-art precision measurements setups based on atom interferometry such as gyroscopes \cite{Dutta2016} or gravimeters \cite{Fang2016,Freier2016} make use of atom sources cooled to the $\mu$K level. In these devices, atom clouds may reach cm-scale size during the interferometric sequence because of thermal expansion, making efficient implementation of cavity-enhanced beam splitters challenging. }
%A simple two-mirror cavity with such a large resonating mode would result in an unstable resonator.

In this paper, we propose to use a marginally stable optical resonator using an intra-cavity focusing element allowing for a resonating mode of several millimeters in diameter. This device therefore enables efficient implementation of cavity-enhanced beam splitters on thermal atom sources. The stability of this cavity critically depends on the relative distance between the optical elements of the resonator. In the context of atom interferometry, longitudinal alignment errors are carefully investigated as they introduce phase distortions of the mode which might impact the sensitivity and accuracy of the interferometer. We first introduce the geometry and the main features of the resonator. We calculate then the intracavity field as a function of alignment errors, carrying out the computations with an iteration code based on the propagation of paraxial rays using the ABCD formalism \cite{Kogelnik1966}. We finally numerically simulate a Bragg atom interferometer with cold atoms to assess the feasibility of LMT atom optics enhanced with this resonator.

\section{A resonator for atom interferometry}\label{reson}

%In the following, we introduce the requirements that will drive the design of a resonator for Atom Interferometry (AI). The two counter counterpropagating cavity fields will be used to manipulated cold atom using two photons transitions and create an interferometer for matter-waves.
In the following, we introduce the requirements for a resonator that will allow for 
%where the cavity stationary field will be used to create an interferometer for matter-waves based on 
$2n$-photons Bragg diffraction of atoms located in the standing wave created by the counter-propagating cavity fields.  Without loss of generality, we will consider as an example that the resonator will be used to interrogate a sample of rubidium atoms at a temperature of $T_e=1 \ \mathrm{\mu K}$ using a set of laser pulses applied up to a time $t_{\text{free}}=250\ \mathrm{ms}$ after preparation. In this time-window, the dispersion in position of the atom cloud is characterized by a Gaussian distribution of maximum 1/$e^2$ width of $t_{\text{free}}\sqrt{k_BT_e/M}=5 \ \mathrm{mm}$\footnote{We assume here that the initial position dispersion of the source is negligible after the time $t_{\text{free}}$.}, where $k_B$ is the Boltzmann constant and $M$ the mass of an atom. In the following, we will therefore take as guideline that the cavity has to be resonant with a fundamental Gaussian beam of waist $\omega_0=5\ \mathrm{mm}$. In order to be easily integrated on common AI experiments \cite{Fang2016,Freier2016} we also set the maximum cavity length to be $L=1\ \mathrm{m}$. As a resonance criteria for this Gaussian beam, we consider that its complex radius of curvature $q_0=iz_{r0}+z_0$ (where $z_{r0}=\frac{\pi\omega_0^2}{\lambda}$ is the Rayleigh length and $z_0$ the waist location) has to remain invariant on a cavity round trip. 
\medskip

If we choose a common symmetric 2-mirror geometry, the only resonators that can match this resonance criteria are:
\begin{itemize}
\item Strictly stable cavities that are resonant with a unique Gaussian beam whose waist size and position are set by the radius of curvature of the cavity mirrors. 
\item Marginally stable confocal cavities that are resonant with any Gaussian beam, given that the cavity mirrors match the confocal criteria.
\end{itemize}
\begin{figure}[htp]
\includegraphics[width=0.4\linewidth]{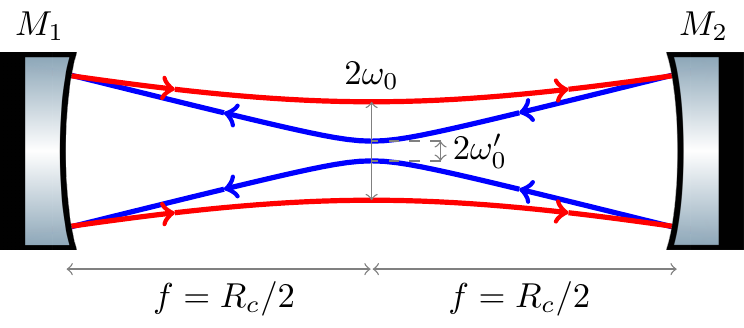}
\centering
\caption{A confocal cavity can be made resonant with a Gaussian beam of waist $\omega_0$=5 mm located at the center of the cavity and propagating in the forward direction (red). The resonating field propagating in the backward direction is then a Gaussian beam of waist $\omega'_0=\frac{\lambda R_c/2}{\pi \omega_0}$ (blue).}
\label{cavity_confocal}
\end{figure}
A strictly stable cavity of length $L$ resonant with a Gaussian beam of waist $\omega_0=5\ \mathrm{mm}$ would have mirrors of radius of curvature of $R_c>20\ \mathrm{km}$, putting extremely stringent requirements on the quality of cavity optics\footnote{The sag of the curved mirror surface for a one inch optics would be smaller than 4 nm which represents a deviation from a perfectly flat surface smaller than $\lambda/200$.}, which rules out this configuration. The marginally stable confocal cavity illustrated Fig.~\ref{cavity_confocal}, for instance with mirrors of radius of curvature $R_c=L$,
%The only possibility is then to use a confocal cavity with mirrors of radius of curvature $R_c=L/2$ as illustrated in Fig.~\ref{cavity_confocal}.
%which is compliant with the resonance condition on a round-trip for a Gaussian beam of waist $\omega_0$=5 mm.
%In this marginally stable configuration, the cavity 
could be made resonant with a Gaussian beam of waist $\omega_0=5\ \mathrm{mm}$ traveling in the forward direction. Nevertheless, the Gaussian beam traveling in the backward direction would be strongly focused with an image waist $\omega'_0=\frac{\lambda R_c/2}{\pi\omega_0}<$ 20 $\mu$m,  which would spoil the atom interrogation process.

These simple arguments show that in order to fulfill the requirements of an atom interferometer, other cavity geometries have to be used.

\subsection{Description of the cavity}

\begin{figure}[h!]
\includegraphics{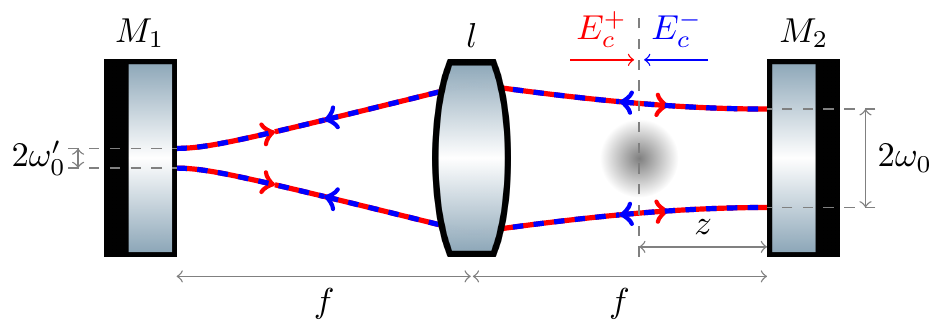}
\centering
\caption{Geometry of the proposed resonator with one lens of focal length $f$ placed at the center of parallel-plane cavity of length $2f$. An input Gaussian beam of waist $\omega'_0$ located on $M_1$ is imaged on $M_2$ with a magnification factor $\lambda f/(\pi\omega_0'^2)$. The atom source is placed at distance z from $M_2$.}
\label{geometry_cavity}
\end{figure}

We propose to use a resonator geometry presented in Fig.~\ref{geometry_cavity}: two flat mirrors $M_1$ and $M_2$ placed at the focal point of an intra-cavity lens $l$.
This resonator has a round-trip ABCD transfer matrix \cite{Siegman1986} of:
\begin{equation}\label{TransferMS}
T=
\begin{pmatrix}
1&f\\
0&1
\end{pmatrix}
\begin{pmatrix}
1&0\\
-1/f&1
\end{pmatrix}
\begin{pmatrix}
1&2f\\
0&1
\end{pmatrix}
\begin{pmatrix}
1&0\\
-1/f&1
\end{pmatrix}
\begin{pmatrix}
1&f\\
0&1
\end{pmatrix}
=
\begin{pmatrix}
-1&0\\
0&-1
\end{pmatrix}=-I
\end{equation}
%the same of the symmetric confocal cavity presented in Fig. \ref{cavity_confocal}. 
meaning that any complex radius of curvature describing a Gaussian beam $q=iz_r+z$ will be transformed back to itself after one round trip inside the cavity. This property is specific to some marginally stable resonators such as the symmetric confocal cavity presented in Fig.~\ref{cavity_confocal}.
%will remain invariant on a cavity round trip and the cavity therefore complies with the resonance criteria previously mentioned. 
%the same of the symmetric confocal cavity presented in Fig. \ref{cavity_confocal} and therefore can be made resonant with any incident Gaussian beams.
Specifically, when the resonator is injected with a Gaussian beam of waist $\omega'_0$ located on the input mirror $M_1$, the lens forms its image waist $\omega_0=\frac{\lambda f}{\pi\omega_0'}$ on $M_2$ which forms back the image of the initial waist $\omega'_0$ on $M_1$.
In comparison with the configuration of Fig.~\ref{cavity_confocal}, this resonator has an unfolded geometry which enables to have identical forward and backwards resonating Gaussian beams. 

We note that a configuration with similar properties has been previously studied in the context of high resolution spectroscopy, in order to obtain large interrogation beams to reduce the transit broadening \cite{Hall1976}. In such work, the focusing element was a parabolic reflector collimating the light coming from its focus on a flat mirror. The use of a lens has the advantage to obtain a resonator with a fully cylindrical symmetry, which is more robust to misalignment.

In the following, we will set $f = 40$ cm and $\omega_0'=20\ \mu$m thus obtaining an image waist of $\omega_0=5\ $mm as required.

\subsection{Calculation of the cavity interrogation fields}\label{interro_fields}

To characterize the standing wave resulting for addition of the counter-propagating cavity fields $E_c^{\pm}(r,z)=|E_c^{\pm}(r,z)|e^{i\varphi^{\pm}(r,z)}$ at position $z$ of the atoms with respect to $M_2$, we 
%During the interrogation process based on Bragg transitions, the atoms located  at a distance $z$ from $M_2$ will be diffracted on the standing wave created by the counter-propagating cavity fields $E_c^{\pm}(r,z)=|E_c^{\pm}(r,z)|e^{i\varphi^{\pm}(r,z)}$ at the position of the atomic ensemble. We now 
calculate the longitudinal and transverse dependence of the product of the fields moduli $|E_c^{+}(r,z)E_c^{-}(r,z)|$ and their phase difference $\Delta\phi(r,z)=\varphi^{+}(r,z)-\varphi^{-}(r,z)$.
\footnote{These terms can respectively limit atom diffraction efficiency and introduce spurious measurement biases. These aspects we be will studied further in Sec.\ref{simuATOM}.}

\begin{figure}[h!]
\centering\includegraphics[width=0.7\linewidth]{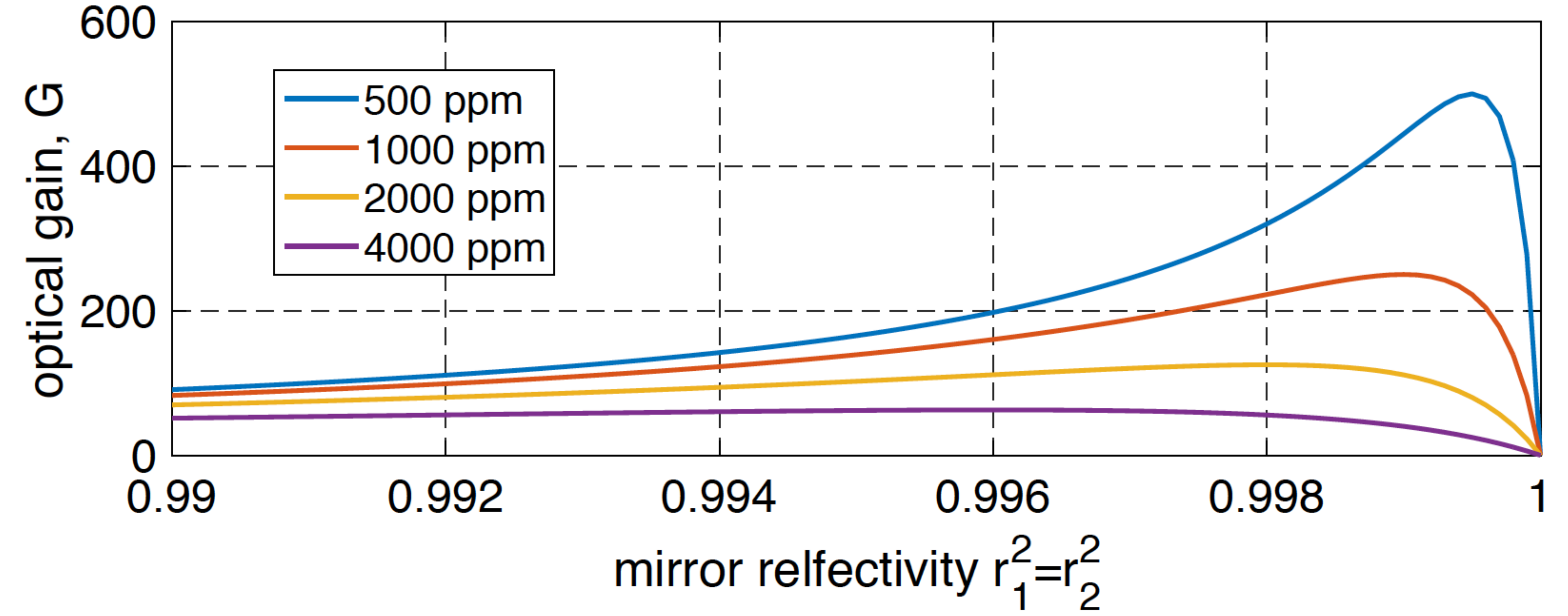}
\caption{Amplification factor of the resonator with respect to the reflectivity of its mirrors for different reflection losses of the intra-cavity lens.}
\label{optical_gain}
\end{figure}

Since $E_c^{\pm}(r,z)$ are Gaussian fields at a distance $\mp z$ from the waist $\omega_0$ located on $M_2$, they may be expressed analytically as:
\begin{equation}
E_c^{\pm}(r,z)=\sqrt{\frac{4\mu_0 cP_c^{\pm}}{\pi}}\frac{1}{\omega(z)}e^{-\frac{r^2}{\omega(z)^2}}
e^{\pm i\phi_{G}(z)} e^{\mp i\frac{kr^2}{2R_c(z)}}e^{\mp ikz}
\end{equation}
where $\omega(z)=\sqrt{\omega_0^2\left[1+\left(\frac{z}{z_{r0}}\right)^2\right]}$, $R_c(z)=z+\frac{z_{r0}^2}{z}$ and $\phi_{G}(z)=\arctan \left(\frac{z}{z_{r0}}\right)$ are respectively the waist size, the radius of curvature and the Gouy phase of the Gaussian beam at the atom position. We also define $\mu_0$ as the vacuum permeability, $c$ as the speed of light in vacuum and $P_c^{\pm}$ as the power of the $E_c^{\pm}$ fields.
We therefore obtain:
\begin{eqnarray}
|E_c^{+}(r,z)E_c^{-}(r,z)|&=&\frac{4\mu_0c G P_{in}}{\pi w(z)^2}e^{-\frac{2r^2}{w(z)^2}}\label{idealfield}\\
\Delta\phi(r,z)&=&-\frac{kr^2}{R_c(z)}+2\phi_{G}(z).\label{idealphase}
\end{eqnarray}
%
%
%\begin{equation}
%\Delta\phi(r,z)=\frac{-2kr^2}{2R_c(z)}+2\phi_{G}(z)
%\end{equation}
%
%
%\begin{figure}[h!]
%\includegraphics[width=0.6\linewidth]{figures/field_article.pdf}
%\centering
%\caption{}
%\label{cavityGain}
%\end{figure}
%
%
Where $G$ is the optical gain of the cavity defined as $G=\sqrt{P_c^+P_c^-}/P_{in}$.
%that will be a key parameter for LMT applications. 
This term can be expressed as a function of cavity parameters: if we note $r_1$, $r_2$, $r_{\mathrm{L}}$ the amplitude reflection coefficients of respectively $M_1$, $M_2$ and $l$ that we consider without losses, $G$ becomes (see \ref{annexA}):
\begin{equation}
G  = \dfrac{r_2(1-r_1^2)(1-r_{\mathrm{L}}^2)}{[1 - r_1 r_2 (1-r_{\mathrm{L}}^2)]^2}. 
\label{Eq_optical_gain}
\end{equation}
For a cavity formed with identical mirrors, Fig.~\ref{optical_gain} shows the amplification factor as a function of squared reflectivities of $M_1$ and $M_2$ for lens reflections $r_{\mathrm{L}}^2$ between 500 ppm and 4000 ppm. We observe that optical gain of a few hundreds can be achieved by using an intra-cavity lens with standard anti-reflection treatments.

In the following, we will set $r_{\mathrm{L}}^2= 1000$~ppm and $r_1^2=r_2^2= 0.9912$. These settings yield an optical gain of $G=100$ and a Finesse of $\mathcal{F}=355$.

%\textcolor{red}{DIRE DEUX MOTS SUR LE GAIN QUI DOIT ETRE FIXE ICI}
%Due to the intra-cavity losses, there is an optimum for the optical gain which does not coincide with the maximum of the finesse.
%With this resonator geometry, it is possible to achieve, with commercial optics, optical gains of a few hundreds. Such amplification factors can be exploited for high order LMT beam splitters, or for operating conventional atom interferometers with reduced input power.

\subsection{Influence of longitudinal misalignments}

%The above treatment to calculate the resonating field of the cavity stands for a lens ($l$) placed at an exact distance $f$ of both mirror ($M_1$) and ($M_2$). This imaging configuration makes the cavity marginally stable thanks to its round trip transfer matrix of -Id (Equ. \ref{TransferMS}). \textcolor{teal}{we repeat what is said in the previous part}

Longitudinal misalignments of the optical elements with respect to the ideal $f-f$ configuration will make the resonator strictly stable or unstable and thus modify its resonating field for the same input Gaussian beam. If the distance between $l$ and $M_{1,2}$ is changed to $f+\delta_{1,2}$, the round-trip ABCD matrix of the cavity becomes:

\begin{equation}
T'=
\begin{pmatrix}
-1+\frac{2\delta_{1}\delta_{2}}{f^2}&2\delta_{1}(-1+\frac{2\delta_{1}\delta_{2}}{f^2})\\
\frac{2\delta_{2}}{f^2}&-1+\frac{2\delta_{1}\delta_{2}}{f^2}
\end{pmatrix}
=
\begin{pmatrix}
A&B\\
C&D
\end{pmatrix}.
\end{equation}
The resonator will be strictly stable if $|A+D|<2$ and unstable if $|A+D|>2$. In other words, the resonator will be strictly stable if $\delta_1$ and $\delta_2$ have the same sign, and unstable if they have opposite signs (the ideal case $\delta_1=\delta_2=0$ corresponding to the marginally stable configuration). 
%\textcolor{teal}{the sentence is a bit heavy}

While calculation of the resonating field in the strictly stable case can be again  evaluated analytically in terms of decomposition of the input beam on the cavity modes, the mathematical description of the resonator in the unstable case is more complex \cite{Siegman1974}.
In the following, we introduce a formalism to calculate the resonating field in all stability conditions and study variation of phase and amplitude of the resonating field as a function of misalignments $\delta_{1,2}$. In practice such misalignments could result from temperature fluctuations, vibrations or imprecisions in the cavity assembling.
We  will show  that the properties of the intra-cavity field will only be slightly modified for the typical misalignments ($\delta_{1,2}\sim 10 \ \mu m$) which might be  experimentally encountered, and that such imperfections are thus not critical for atom interferometry applications.

\subsubsection{Calculation of the interrogation field taking into account longitudinal misalignments.}

We use an iterative approach to calculate the field resonating in the cavity at position of the atoms. Our method is based on ABCD-matrix propagation of the input Gaussian beam through its round-trip inside the resonator.
In the following, we note $q_n^{\pm}$ the complex radius of curvature of the Gaussian beam at position of the atoms in the forward (+) and backwards (-) directions after $n$ round trips inside the cavity.
\begin{figure}[h!]
\centering
\includegraphics[scale=0.8]{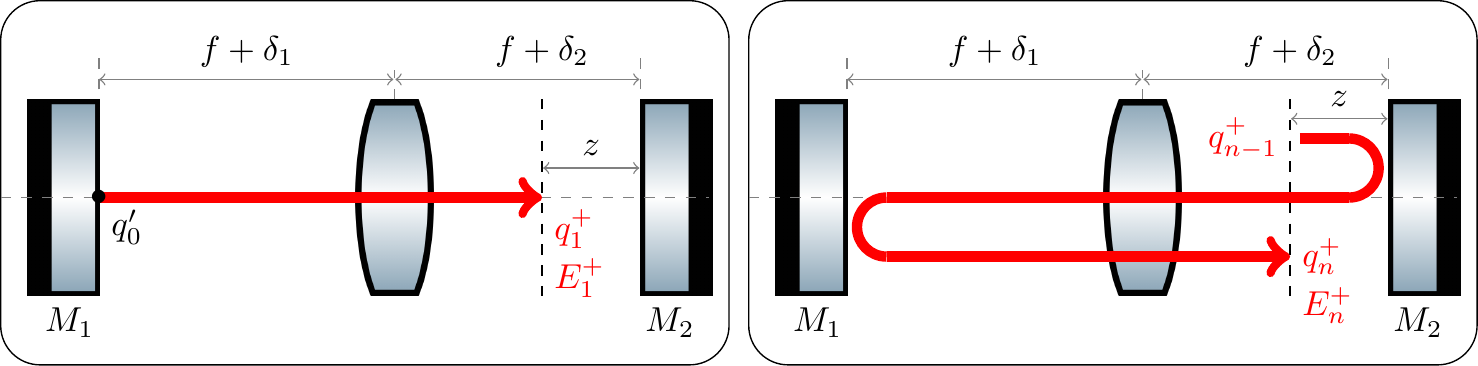}
\caption{Determination of the $q^+$ parameter for the field propagating to the right: for the first round trip (left), and for the n-th round trip with $n>1$ (right).}
\label{qplus}
\end{figure}
For the first round trip, $q_1^{+}$ is obtained from the input beam $q'_0 = i\frac{\pi (\omega'_{0})^2}{\lambda}$ by using the  ABCD transfer matrix (see Fig.~\ref{qplus}-left):
\begin{equation}
U=
\begin{pmatrix}
1&\delta_{2}+f-z\\
0&1
\end{pmatrix}
\begin{pmatrix}
1&0\\
-1/f&1
\end{pmatrix}
\begin{pmatrix}
1&\delta_{1}+f\\
0&1
\end{pmatrix}
\end{equation}
Then for the n-th round trip with n$>$1, the complex radius of curvature $q_n^{+}$ is obtained from $q_{n-1}^{+}$ using the transfer matrix (see Fig.~\ref{qplus}-right):
\begin{equation}
V=
\begin{pmatrix}
1&\delta_{2}+f-z\\
0&1
\end{pmatrix}
\begin{pmatrix}
1&0\\
-1/f&1
\end{pmatrix}
\begin{pmatrix}
1&2(\delta_{1}+f)\\
0&1
\end{pmatrix}
\begin{pmatrix}
1&0\\
-1/f&1
\end{pmatrix}
\begin{pmatrix}
1&z+\delta_{2}+f\\
0&1
\end{pmatrix}
\end{equation}
\begin{figure}[h!]
\centering
\includegraphics[scale=0.8]{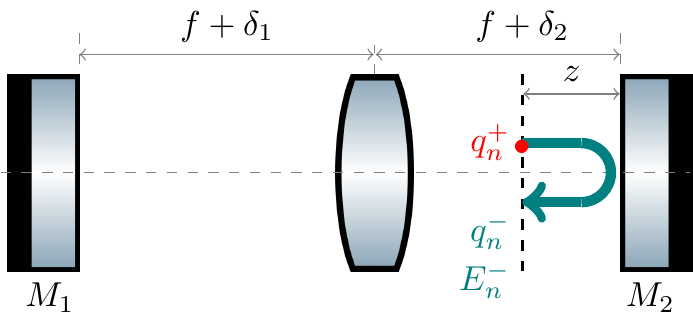}
\caption{Determination of the $q^-$ parameter for the field propagating to the left.}
\label{qmoins}
\end{figure}
For all round trips, the complex radius of curvature of the beam propagating in the backwards direction $q_n^{-}$ can be obtained from $q_{n}^{+}$ using the transfer matrix (see Fig.~\ref{qmoins}):
\begin{equation}
W=
\begin{pmatrix}
1&2z\\
0&1
\end{pmatrix}.
\end{equation}
We therefore obtain the following iterative expression for $q_n^{\pm}$:
\begin{equation}
\left \{
\begin{array}{c c l l}
q_1^{+}  &=& \frac{U_{11}.q_0+U_{12}}{U_{21}.q_0+U_{22}} &  \\
 & & \\
q_{n}^{+}  &=& \frac{V_{11}.q_{n-1}^{+}+V_{12}}{V_{21}.q_{n-1}^{+}+V_{22}} & \textrm{ for } n> 1 \\
 & & \\
q_{n}^{-}  &=& \frac{W_{11}.q_{n}^{+} +W_{12}}{W_{21}.q_{n}^{+} +W_{22}} & \textrm{ for } n\geqslant 1.
\end{array}
\right.
\end{equation}
We obtain the expressions for the beam size $\omega_{n}^{\pm}$ and the real radius of curvature $R_{n}^{\pm}$ of the Gaussian beam at position of the atoms in the forward (+) and backwards (-) directions after n round trips inside the cavity:
\begin{equation}
\left \{
\begin{array}{c c l}
\omega_{n}^{\pm}&=&\sqrt[]{\frac{-\lambda}{\pi \mathrm{Im}(1/q_n^{\pm})}}  \\
R_{n}^{\pm}&=&\frac{1}{\mathrm{Re}(1/q_n^{\pm})} 
\end{array}
\right.
\end{equation}
The accumulated Gouy phase shift $\phi_{G,n}^{\mathrm{acc}\pm}$ can be calculated in an iterative way \cite{Erden1997} considering that the phase reference is taken in the plane of $q_1^{+}$, ie. $\phi_{G,1}^{\mathrm{acc}+}=0$
\begin{equation}
\left \{
\begin{array}{c c l l}
\phi_{G,1}^{\mathrm{acc}+}&=&0 & \\
\phi_{G,n}^{\mathrm{acc}+} &=& \phi_{G,n-1}^{\mathrm{acc}+} + \tan^{-1}\left( \frac{\lambda V_{12}}{\left(V_{11}+\frac{V_{12}}{R_{n-1}^+}\right)\pi (\omega_{n-1}^+)^2} \right) & \textrm{ for n}>1 \\
\phi_{G,n}^{\mathrm{acc}-} &=& \phi_{G,n-1}^{\mathrm{acc}-} + \tan^{-1}\left( \frac{\lambda W_{12}}{\left(V_{11}+\frac{W_{12}}{R_{n-1}^-}\right)\pi (\omega_{n-1}^-)^2} \right) & \textrm{ for n}\geqslant 1 
\end{array}
\right.
\end{equation}
The Gaussian field $E_n^{\pm}(r,z)$ at the position of the atom after n round trips can then be expressed:   
\begin{equation}
\left \{
\begin{array}{c c l}
E_n^+(r,z) &= &t_1t_L(r_1 r_2 t_{\mathrm{L}}^2)^{n-1}\sqrt{\frac{4\mu_0 cP_{\mathrm{in}}}{\pi}}\frac{1}{\omega_n^+}e^{-\frac{r^2}{(\omega_n^+)^2}}
e^{-2ikn L}e^{i\phi_{G,n}^{\mathrm{acc}+}} e^{-i\frac{kr^2}{2R_n^+}}
  \\
E_n^-(r,z) &= &t_1t_Lr_2(r_1 r_2 t_{\mathrm{L}}^2)^{n-1}\sqrt{\frac{4\mu_0 cP_{\mathrm{in}}}{\pi}}\frac{1}{\omega_n^-}e^{-\frac{r^2}{(\omega_n^-)^2}}
e^{-2ik(nL+z)}e^{i\phi_{G,n}^{\mathrm{acc}-}} e^{-i\frac{kr^2}{2R_n^-}} 
\end{array}
\right.
\end{equation}
where $L=2f+\delta_{1}+\delta_{2}$ is the length of the cavity and $k=2\pi/\lambda$ is the wave number.
The total intra-cavity field $E_c^\pm(r,z)$ is the sum of the fields $E_n^\pm(r,z)$: 
\begin{equation}\label{champTOT}
E_c^\pm(r,z) = \sum_{n=1}^N E_n^\pm(r,z). 
\end{equation}

\subsubsection{Study of the influence of misalignments.}
We now use the above method to calculate numerically $|E_c^{+}(r,z)E_c^{-}(r,z)|$ and $\Delta\phi(r,z)$ for different longitudinal misalignments $\delta_{1,2}$. For each value of $\delta_{1,2}$, we calculate numerically the sum of Eq.~\eqref{champTOT} maximizing the power in transmission of the cavity by changing the input wavelength. This last criteria ensures that the cavity is locked correctly (see \ref{annexB} for details of the simulation).

\begin{figure}[h!]
\centering
\includegraphics[width=0.9\linewidth]{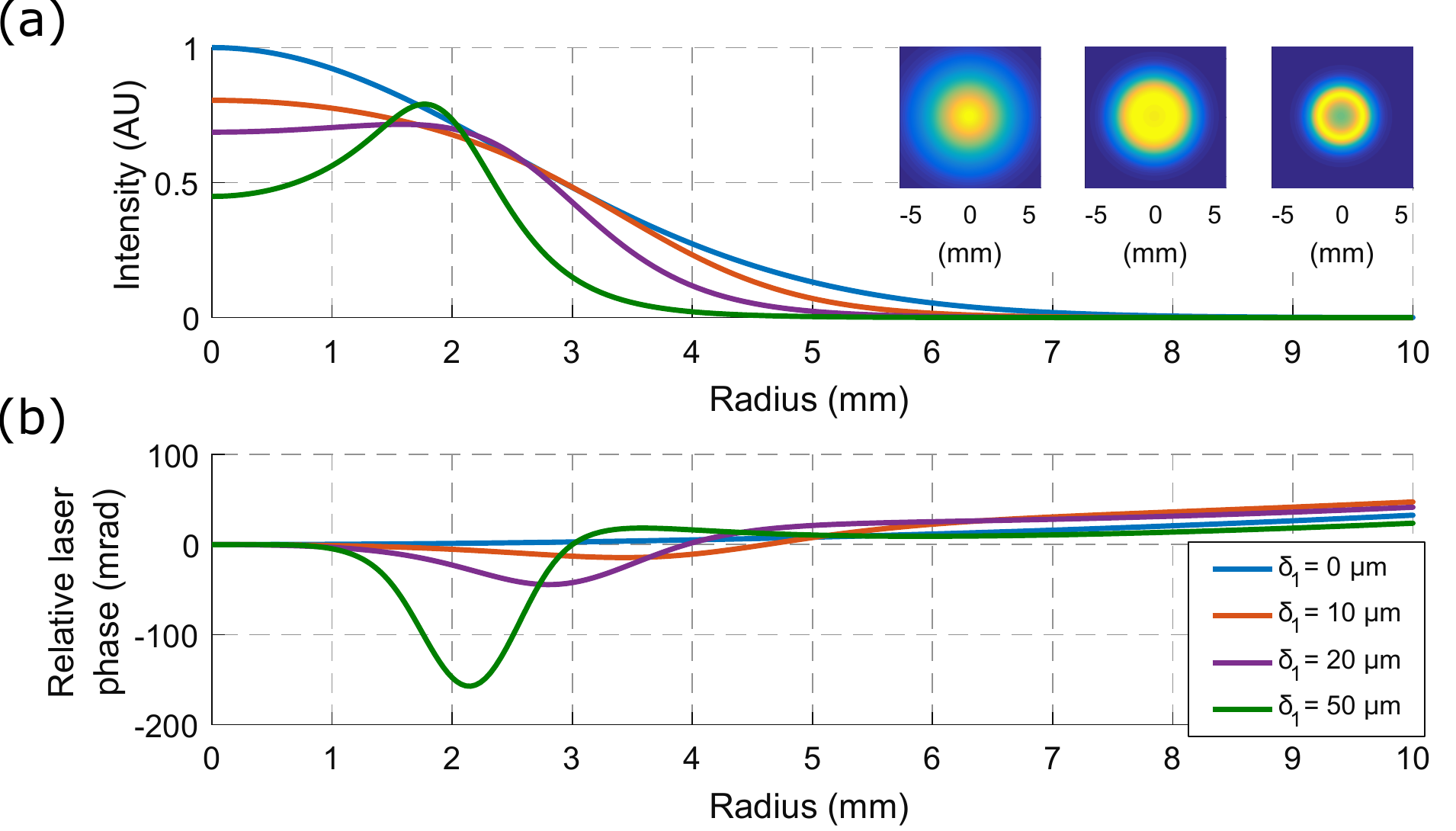}
\caption{(a) Radial intensity profile of the cavity field $E_c^+(r,z)$  at the middle point $z$ between $l$ and $M_2$, for various misalignments  $\delta_1=0, \ 10, \ 20, \ 50 \ \mu$m. 
Inset: 2D plot of the corresponding normalized intensity cross-section for  $0, \ 20, \ 50 \ \mu$m.
(b) Radial profile of the relative phase $\Delta\phi(r,z)$ between the two counter-propagating cavity  fields $E_c^+(r,z)$ and $E_c^-(r,z)$.}
\label{field_shape_misaligned}
\end{figure}

Fig.~\ref{field_shape_misaligned} shows the results of the calculation for various displacements $\delta_1$ of the input mirror with respect to the ideal configuration while keeping $\delta_2=0$.
Panel (a) shows the field intensity profiles of the two counter-propagating cavity fields at the atom position, and panel (b) their relative phase.
 When the resonator is perfectly aligned ($\delta_1=\delta_2=0$), the fields are Gaussian with an almost uniform relative phase. In this ideal alignment configuration, we verified that the results of the numerical calculation give the same results as the analytic expressions of Eq.~\eqref{idealfield} and \eqref{idealphase}.
The displacement of the input mirror determines a ring-shaped intensity distribution of the cavity fields, and increases the inhomogeneity  of their relative phase. These inhomogeneities are due to the fact that the fields do not perfectly overlap  after each round-trip and acquire a spatially-dependent phase shift. Moreover, misalignments degrade the optical gain that does not reach anymore the maximum value of $G=100$ given by Eq.~\ref{Eq_optical_gain}. More precisely, the optical gain equals $100, \ 80, \ 66, \ 42$ for $\delta_1=0, \ 10, \ 20, \ 50 \ \mu$m, respectively. The drop of the gain results from the progressive dephasing of the fields after each round trip.

Finally, displacements $\delta_2$ of the back mirror $M_2$ have little impact on the intensity and phase profile of the intra-cavity fields compared to the displacements of $M_1$ (see \ref{annexC}). Indeed, close to $M_2$ the fields are nearly plane waves and variations of $\delta_2$ therefore have small impact on their transverse phase dependence.

%In the next section, we will discuss the influence of the distortions of the intra-cavity fields on the performances of the atom interferometer.

\section{Large momentum transfer atom interferometry}\label{simuATOM}
We discuss now the influence of the distortions of the intra-cavity fields on the performances of the atom interferometer.
\subsection{Configuration}
%We consider an atom interferometer that exploits the optical gain of our resonator to enhance its sensitivity via LMT techniques. If the interferometer is operated on ground, the atomic ensemble moves with respect to the cavity setup because of the gravitational acceleration. Two configurations can be adopted. First, a vertical interrogation cavity with the atoms moving along its axis, as in \cite{Hamilton2015}. 
%This solution requires the chirp of the cavity length to keep the interrogation beams resonant with the freely falling atoms, for example acting on the position of one cavity mirror \cite{Briles2010}. 
%A systematic effect to be considered in this case is the different laser phases experienced by  the atoms along the cavity axis during their displacement. 
%A second configuration can be adopted, based instead on a horizontal interrogation \cite{Dutta2016}. The  motion of the atomic clouds along the gravity axis imposes in this case an interrogation cavity at each position where the atoms are manipulated. 
%In the case of a  $\pi/2-\pi-\pi/2$ light pulse sequence with the $\pi$ pulse  at the  apogee of the trajectory (as in \cite{Canuel2016}), two vertically separated horizontal cavities are required. The two interrogation cavities must be aligned and stabilized with high precision to control systematic effects and to maintain the fringe visibility \cite{Dutta2016}.

\begin{figure}[h!]
\centering
\includegraphics[width=0.7\linewidth]{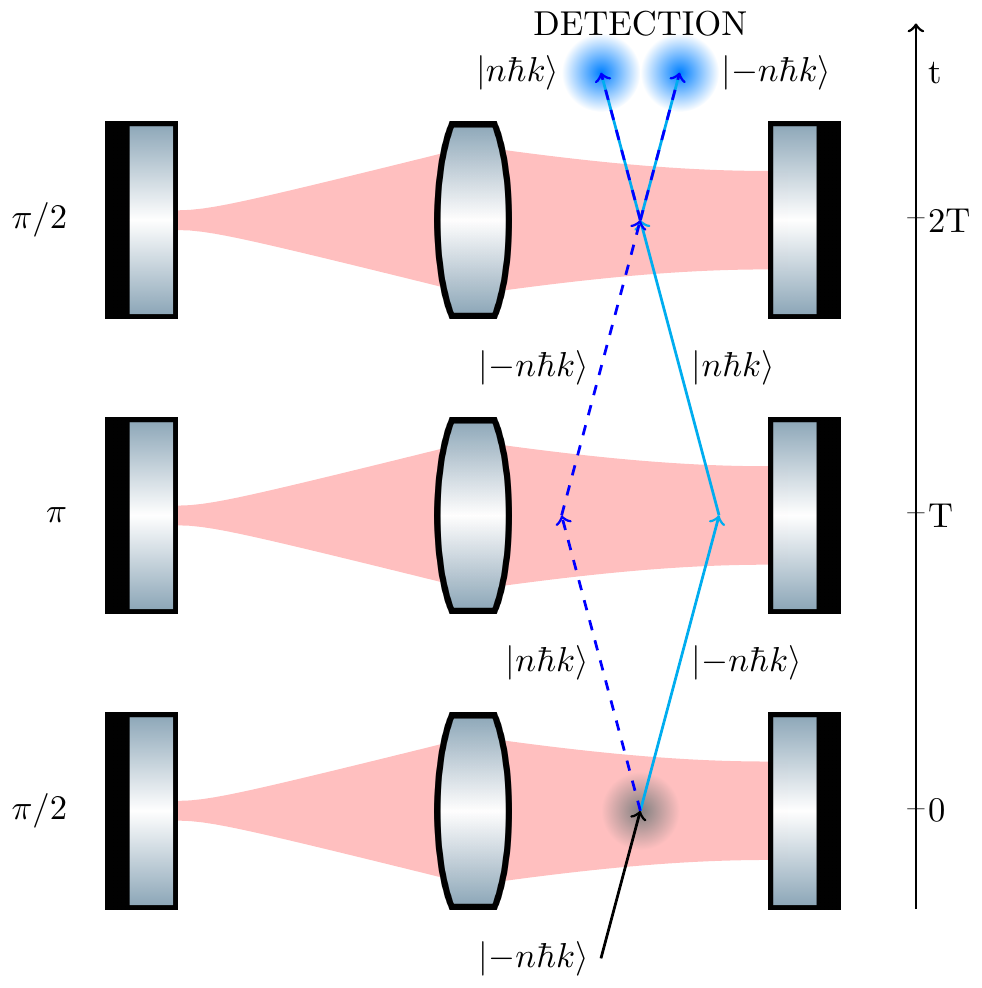}
\caption{Cavity enhanced atomic accelerometer operated in the absence of external acceleration. The atomic ensemble, initially  at the mid point between $l$ and $M_2$ and with momentum $\ket{-n \hbar k}$, is split, deflected and recombined with a set of $\pi/2-\pi-\pi/2$ cavity enhanced pulses.}
%\textcolor{red}{Labels too big, esp. $\pi/2 - \pi - \pi/2$ and 0-T-2T.}
\label{fig:microgravAI}
\end{figure}
We study here the simple case of an atom accelerometer using cavity enhanced diffraction in the absence of external inertial forces. In this configuration, the same beam manipulates the atoms with a three pulse sequence modulated in time (see Fig.~\ref{fig:microgravAI}).

%For the sake of simplicity, we study an horizontal accelerometer configuration in microgravity \cite {Geiger2011} so that the same beam coherently manipulates the atoms with a three pulses sequence. The configuration is  shown in Fig.~\ref{fig:microgravAI}.  

The atomic ensemble considered here has a radial temperature of  $T_e=1\ \mu$K. The width of the velocity distribution along the interrogation direction is set to be much smaller than the recoil velocity so as to avoid the spatial overlap of the two interferometer output ports after the interferometer. This condition can be obtained with a long velocity selection  pulse prior to the interferometric sequence. 
The same pulse can be used to fulfill the order $n$ Bragg reflection condition by setting the initial velocity of the ensemble to $v_0=-n\hbar k/M$, where $k=\omega/c$ is the wavenumber of the single laser injected in the cavity to create the standing wave.
%The same pulse can be used to set the initial velocity $v_0=-n\hbar k/M$ of the ensemble to fulfill the order $n$ Bragg reflection condition. A single laser, of wavenumber k, is injected in the cavity to create the standing wave.
%with the standing wave created in the cavity injected with a single laser  of wavevector $\mathbf{k}$.
 The position of the cloud at the time of the first interrogation pulse is set at the midpoint between $l$ and $M_2$.
%Ideally, the cavity standing wave couples atomic states with momentum $\mathbf{p}=\pm n\hbar\mathbf{k}$, and generates Rabi oscillations between these two external states. 
%However, different diffraction orders might be excited, because of the finite pulse length (i.e. the Rabi frequency of finite width) with the related frequency windowing, and the finite radial temperature of the atomic ensemble, which determines its increase in size and in turn its probing of the radial profile of the interrogation beams.

\subsection{Simulation}
To numerically evaluate the effects of the fields inhomogeneities, we consider an ensemble of atoms as two level systems with ground state $\ket{g}$ and  excited state $\ket{e}$ (energy separation $\hbar \omega_{eg}$) interacting with two counter-propagating beams with frequency $\omega$.
%We define $k=2\pi/\lambda=\omega/c$ the optical wavevector of the laser, and $\Delta=\omega_{eg}-\omega$ the laser detuning with respect to the optical transition. 
The laser detuning $\Delta=\omega_{eg}-\omega$ is assumed to be much larger than the linewidth of the excited state. In the Rotating Wave Approximation, and after adiabatic elimination of the excited state $\ket{e}$, the evolution of the state $ \ket{\Psi(t)} = \sum\limits_{m} a_m(t) \ket{m}$, where $\ket{m} \equiv \ket{g, m \hbar \mathbf{k}}$, in the laser field is given by a tri-diagonal Hamiltonian matrix $H$ (see, e.g. Ref.~\cite{Muller2008a}), with the diagonal terms given by the atom kinetic energy:
\begin{equation}
\bra{m} H(t) \ket{m} = - m^2\hbar\omega_r
\end{equation}
and the off-diagonal elements representing the coupling between the momentum states:
\begin{equation}
\bra{m \pm 2} H(t) \ket{m} = \frac{\hbar\Omega(t)}{2} e^{\pm i \Delta \phi}.
\end{equation}
 Here $\omega_r = \hbar k^2/2M$ is the single photon recoil frequency, and $\Omega \propto |E_c^+||E_c^-|$  the time-dependent two photon Rabi frequency. 
We evaluate numerically the evolution of the state $\ket{\Psi(t)}$ in the case of Gaussian pulses $\Omega(t) = \bar{\Omega} e^{-t^2/2\sigma^2}$, where $\bar{\Omega}$ is the peak Rabi frequency, since they have been shown to enhance the efficiency of LMT beam splitters \cite{Muller2008a,Giese2013}.

\begin{figure}[h!]
\centering
\includegraphics[width=0.8\linewidth]{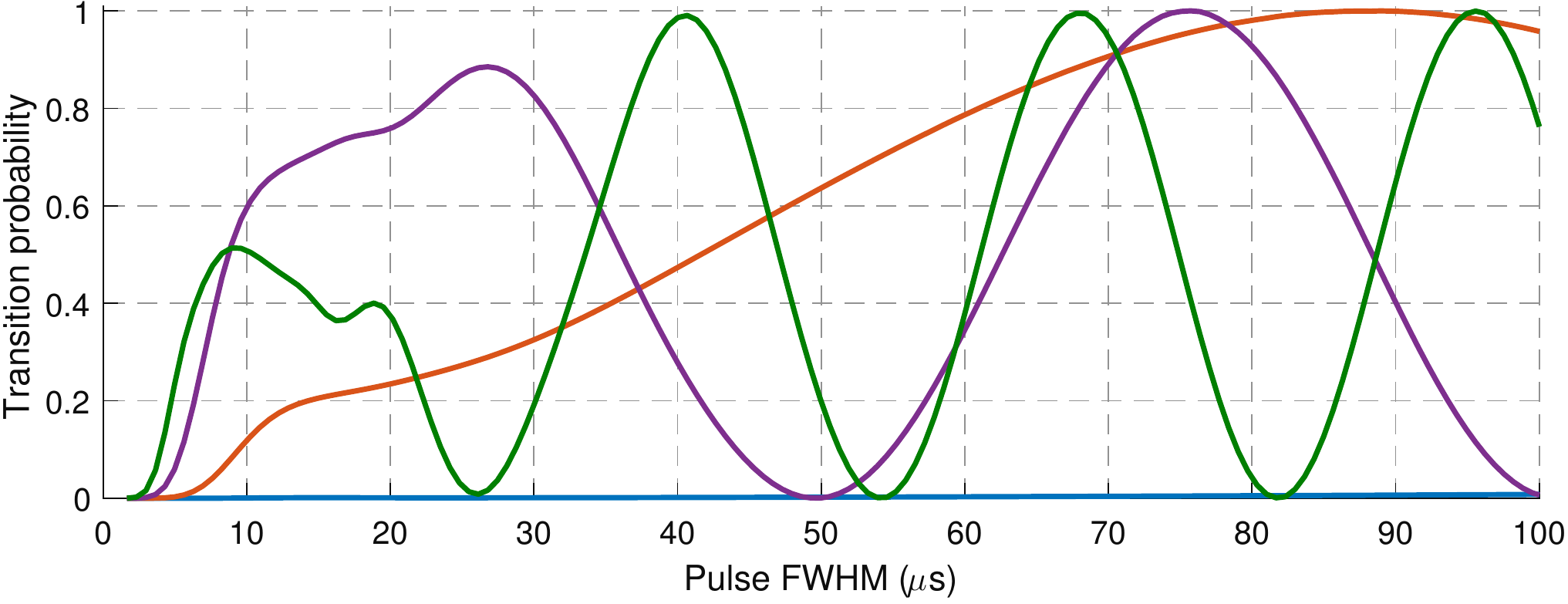}
\caption{
LMT Rabi oscillation for four different peak Rabi frequencies from state $\ket{-5}$ to $\ket{+5}$. We simulate gaussian pulses of varying duration, characterized by their Full Width Half Maximum (FWHM). The peak frequencies are $50$ kHz (blue), $100$ kHz (red), $150$ kHz (purple) and $200$ kHz (green).}
\label{coupling_efficiency_LMT_perfect_case}
\end{figure}

\bigskip
%\subsubsection{Zero temperature.}
%We first consider the case of an atom at zero temperature interacting with a spatially homogeneous laser field, to recall that LMT Bragg diffraction requires large Rabi frequencies and show the advantage in the use of a cavity. 
We first neglect the velocity distribution of the atomic sample. In Fig.~\ref{coupling_efficiency_LMT_perfect_case} we plot the transfer efficiency from the $\ket{-5}$ state to the $\ket{+5}$ state for different values of $\bar{\Omega}$, as a function of the pulse duration $\delta t_{\textrm{FWHM}}$. 
The numerical integration of the Schrödinger equation was performed by restricting the dimension of the Hamiltonian matrix  to the momentum states  $\ket{-9}, \ket{-7}, ...,\ket{+7}, \ket{+9}$.
For each pulse duration, the transfer efficiency   from the initial state was obtained by  integration  on the time interval $\left[ -4 \sigma, 4 \sigma \right]$. We verified that considering more momentum states or integrating over a larger time interval did not modify the results.
For low Rabi frequencies ($<50 \ \text{kHz}$), no appreciable transfer is obtained to the target state (blue curve). Increasing the frequency (100 kHz - red curve) starts to couple the different momentum states. Further increasing the frequency  (150 kHz - purple curve and 200 kHz - green curve) brings to the Bragg regime, where a Rabi oscillation between the two momentum states $\ket{\pm 5}$ is observed. 
High diffraction efficiencies in the Bragg regime are thus obtained when $\bar{\Omega} \times \delta t_{\textrm{FWHM}} \simeq $1.
These results are in agreement with previous studies (see, e.g. Fig. 7 of \cite{Muller2008a}). 

To illustrate the advantage of the cavity build up, we calculate the required input laser power to achieve a Rabi frequency of 200 kHz. We consider  atoms  in the $|F=1, m_F = 0\rangle$ state and  $\sigma^+$ polarized light. The coupled states are therefore $F^\prime=1,2$ and the effective Rabi frequency of the Bragg transition is
\begin{equation}
\bar{\Omega} = \frac{\Gamma^2}{2}\sqrt{\frac{I_c^+ I_c^-}{4I_{sat}^2}}\left(\frac{5}{24\Delta} + \frac{3}{24(\Delta + \Delta_2)}\right)
\end{equation}
with $\Gamma\simeq 2\pi\times 6.06 \ \text{MHz}$ the excited state linewidth, $I_{sat}=1.67 \ \mathrm{mW.cm}^{-2}$ the saturation intensity, $I_c^\pm$ the intensity of the two counter propagating cavity fields, $\Delta$ the detuning from the first coupled state ($F^\prime=1$) and $\Delta_2 = 2\pi\times 157 \ \text{MHz}$ the splitting between $F^\prime = 1$ and $F^\prime = 2$ \cite{Steck2001}.
%Using the relation between the intensity and the power for a Gaussian beam of waist $\omega_0$, and 
Introducing the input power $P_{in}$ and the cavity amplification factor $G$, this expression can be re-written as
\begin{equation}
\bar{\Omega} = \frac{\Gamma^2}{2I_{sat}}\frac{G P_{in}}{\pi\omega_0^2}\left(\frac{5}{24\Delta} + \frac{3}{24(\Delta + \Delta_2)}\right).
\end{equation}
Considering a  detuning $\Delta=2\pi\times 5 \ \text{GHz}$ and a cavity waist $\omega_0=5$~mm, a peak Rabi frequency of 200 kHz can then be achieved with an amplification factor $G=100$ and an input power of $P_{in}= 2.2$~mW.

\bigskip
%\subsubsection{Finite transverse temperature.}
We now investigate the combined effect of the transverse velocity distribution of the atoms and the transverse profile of the interrogation laser beam on the Bragg diffraction efficiency. 
Thanks to the use of a long velocity selection pulse prior to the interferometric
sequence, we consider that in the longitudinal direction, the velocity distribution is sufficiently narrow so as to neglect Doppler effect.
%The expansion of the atomic cloud in the radial directions with respect to the interrogation beams is determined by the temperature $T_e$. The velocity distribution along the interrogation direction is considered sufficiently narrow so as to neglect Doppler effect. This condition is implemented experimentally by the use of a long velocity selection Bragg pulse before the interferometric sequence.

\begin{figure}[h!]
\centering
\includegraphics[width=0.9\linewidth]{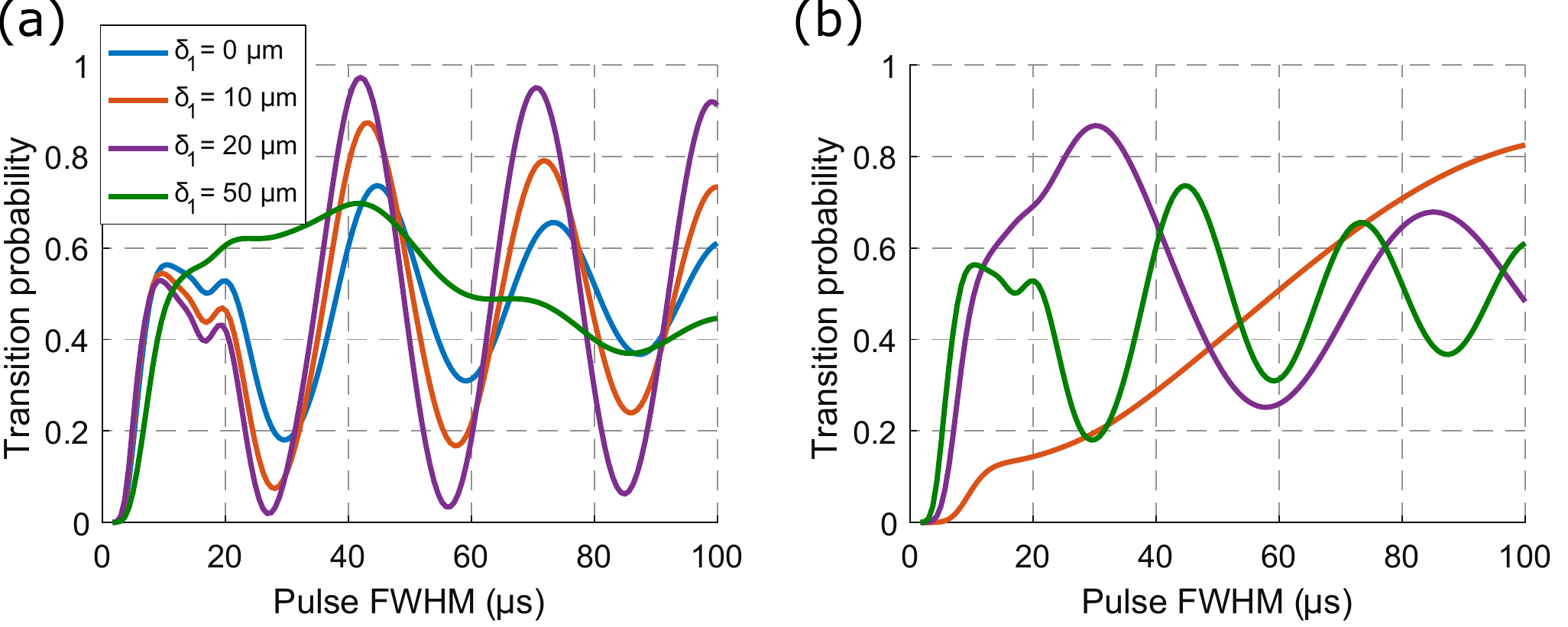}
\caption{Rabi oscillation from the $\ket{-5}$ state to the $\ket{+5}$ state for an atom cloud with 1 $\mu$K transverse temperature, after 100 ms propagation time. (a) The different curves correspond to different longitudinal misalignments of $M_1$ with the peak Rabi frequency kept constant at 200 kHz. (b) The different curves correspond to different peak Rabi frequencies 100 kHz (red), 150 kHz (purple) and 200 kHz (green) in the case of a perfectly aligned cavity.}
\label{mean_rabi_oscillation}
\end{figure}

In Fig.~\ref{mean_rabi_oscillation} the coupling efficiency from the initial $\ket{-5}$ state to the $\ket{+5}$ state is studied for a pulse of variable duration and applied on the atomic ensemble after a free evolution of 100 ms, for different longitudinal misalignments $\delta_1$ of mirror $M_1$ (Fig.~\ref{mean_rabi_oscillation}(a)) and different peak Rabi frequencies (Fig.~\ref{mean_rabi_oscillation}(b)).
The results show that a transfer efficiency close to 100\% is achievable in this resonator for $\delta_1 \leq 20\ \mathrm{\mu m}$. The Rabi oscillations even improve when the cavity is slightly misaligned (orange and violet curves in Fig.~\ref{mean_rabi_oscillation}(a)) because the intensity becomes more homogeneous than for a Gaussian beam (see Fig.~\ref{field_shape_misaligned}). For larger misalignments, the important spatial inhomogeneity strongly degrades the Rabi oscillations. The simulations of Fig.~\ref{mean_rabi_oscillation}(a) were performed by keeping a constant peak Rabi frequency equal to 200 kHz, although the optical gain decreases when the cavity is misaligned. The effect of reducing the Rabi frequency in a perfectly aligned cavity (Gaussian beam probed by the thermal atoms) is shown in Fig.~\ref{mean_rabi_oscillation}(b) for comparison. 

\bigskip
We now consider the $\pi/2-\pi-\pi/2$ cavity-enhanced pulse sequence shown in Fig.~\ref{fig:microgravAI} and choose a total interrogation time $2T=200\ \mathrm{ms}$. During each Bragg diffraction process of order $n$, the relative laser phase is imprinted $n$ times on the matter-waves; the final phase shift for the whole interferometric sequence is thus 
\begin{equation}\label{PhiAtTot}
\Delta\Phi = n\times \left( \Delta\phi_1 - 2\Delta \phi_2+\Delta \phi_3 \right)
\end{equation}
where $\Delta \phi_i$ is the relative phase of the lasers at the position of the atom during the $i$-th laser pulse \cite{Kasevich1991}. The sensitivity to the inhomogeneities of the relative laser phase is thus $n$ times higher. This could be a critical issue in relation to the longitudinal misalignment of the cavity elements, which modify the relative phase profile of the counter-propagating fields as presented in Fig.~\ref{field_shape_misaligned}.

\begin{figure}[htp!]
\centering
\includegraphics[width=1\linewidth]{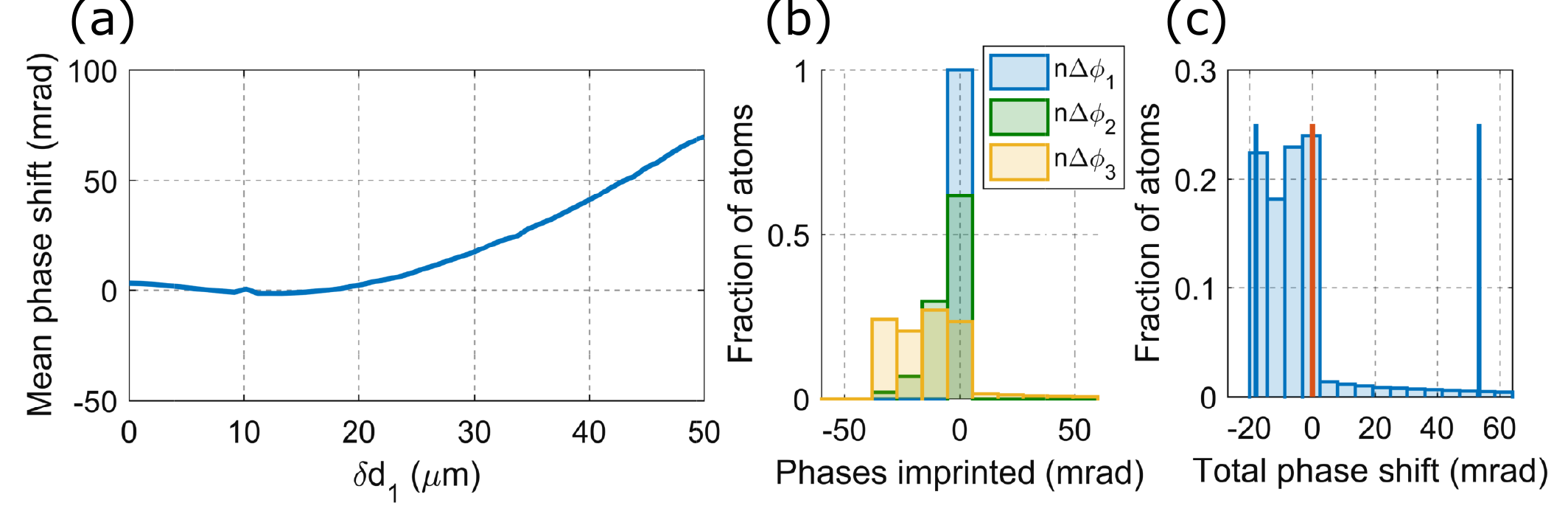}
\caption{
(a) Mean interferometer phase shift associated with the inhomogeneity of the relative  phase of the counter propagating cavity fields  caused by a misalignment of the cavity and the spread of the 1 $\mu$K atom cloud between the three interrogation pulses of a 200 ms interferometer. 
(b) Histograms of the phases imprinted on the atoms at the three pulses for a $\delta_1 = 10 \ \mu$m input mirror misalignment.
(c) Histogram of the total interferometer phase shift for $\delta_1 = 10 \ \mu$m. The red line shows the mean phase shift and the blue lines show the 5\% and 95\% percentiles. Histograms (b) and (c) are plotted considering 500 000 random trajectories.
}
\label{histogramme_phases}
\end{figure}

%For $\delta_1$=10 $\mu$m, the relative phase of the counter-propagating beams in the cavity is shown in Fig.~\ref{histogramme_phases}(a). 
To evaluate the interferometric phase shift $\Delta \Phi$ in this configuration, we calculate the trajectory of  each atom sampled out of a Gaussian thermal distribution of velocities corresponding to the  temperature of 1 $\mu$K, and deduce the position-dependent phase shift. 
Fig.~\ref{histogramme_phases}(a) shows the mean interferometer phase shift as a function of the input mirror misalignement.
Fig.~\ref{histogramme_phases}(b)  shows the histograms of the phases imprinted on the atoms at the three Bragg pulses of the interferometer for the particular value $\delta_1=10\ \mathrm{\mu m}$, and Fig.~\ref{histogramme_phases}(c) the histogram of the total phase shift $\Delta\Phi$ at the end of the interferometer  calculated with Eq.~\eqref{PhiAtTot}. 
Fig.~\ref{histogramme_phases}(a) shows that misalignements of the input mirror induce systematic phase shifts of few mrad only for $\delta_1$ up to $20 \ \mu$m. Such systematic shifts can be characterized by changing the position of the input mirror or the temperature of the atoms. They are therefore not problematic for high precision interferometry. The spread of the phase shifts (e.g. 21 mrad rms for $\delta_1=10\ \mu$m) would lead to a systematic contrast reduction but this effect is negligible for misalignements $\delta_1$ below $20 \ \mu$m. In our simulation, only the transverse variation of the relative laser phase is considered, and not the fact that atoms exploring the edge of the beam -- where the Rabi frequency is lower -- contribute less to the interferometer signal due to reduced transition probabilities. The calculated spread of the phase shift thus represents an upper bound and so the possible contrast reduction.

In conclusion,   longitudinal misalignements    do not significantly affect both the contrast and phase of the interferometer, making this  resonator geometry compatible with precision interferometry based on LMT Bragg pulses.

\section{Conclusion}

We propose a marginally stable optical resonator suitable for  atom interferometry applications. The resonator consists of two flat mirrors placed symmetrically at the focal planes of a lens and magnifies an input waist $\omega'_0$ by a factor $\lambda f/(\pi\omega_0'^2)$.
Waists of several millimeters in the second half of the resonator can be obtained with input beams of $\omega'_0\simeq 20$ $\mu$m and focal length $f\simeq 40 \ \text{cm}$.
Optical gains of several hundreds can be obtained for a lens with losses of about 0.1\% and standard high reflectivity mirrors. Such large beams are suitable for atom interferometry devices such as accelerometers, gradiometers or gyroscopes operating  with  laser cooled atom sources ($T_e\sim \mu$K).
The power enhancement at resonance enables to operate conventional or LMT atom interferometers  with  lower requirements on the input laser power. We investigated the robustness of the resonator  to  longitudinal misalignments in terms of phase and intensity profiles of the intra-cavity field, and illustrated its application to an atom interferometer based on $2\times 5 \hbar k$ LMT Bragg diffraction.
High order LMT in the resonator could for example allow  to operate cold atom inertial sensors at reduced interrogation times in order to increase their bandwidth \cite{Rakholia2014} and reduce their volume.

Future works should focus on the effect of tilts of the optical elements, which can be simulated using  generalized beam matrices \cite{Tovar1995}, and on imperfections of the optical surfaces. Regarding LMT atom optics, the effect of diffraction phase shifts \cite{Buechner2003} and their control in this resonator geometry should be investigated.

\section*{Acknowledgments}
This work was realized with the financial support of the French State through the ``Agence Nationale de la Recherche" (ANR) in the frame of the ``Investissement d'avenir" programs: Equipex MIGA (ANR-11-EQPX-0028), IdEx Bordeaux - LAPHIA (ANR-10-IDEX-03-02) and FIRST-TF (ANR-10-LABX-48-01). This work was also supported by the Région d’Aquitaine (project IASIG-3D) and by the city of Paris (Emergence project HSENS-MWGRAV). We also thank the ``Pôle de compétitivité Route des lasers - Bordeaux'' cluster for his support. G. L. thanks DGA for financial support. M. P. also thanks LAPHIA--IdEx Bordeaux for partial financial support. We thank Bess Fang and Ranjita Sapam for fruitful discussions and experimental work related to this paper.

%We acknowledge financial support from the MIGA Equipex funded by the French National Research Agency (ANR-11-EQPX-0028), from the City of Paris (project HSENS-MWGRAV) and from FIRST-TF (ANR-10-LABX-48-01), from the R{\'e}gion d'Aquitaine (project IASIG-3D), the cluster LAPHIA (project MIGA-PHYS). We thank the ``Pôle de compétitivité Route des lasers - Bordeaux'' cluster for his support.

\section*{References}
\bibliographystyle{unsrt}
\bibliography{large_mode_resonator}

\appendix

%\section{Appendix}

\section{Calculation of the cavity amplification factor G.}\label{annexA}

We note in the following $E_{\mathrm{in}}$ the incident field on the cavity, $E_{cM_1}$ the stationary intra cavity field of power $P_{cM_1}$ propagating in the forward direction on mirror $M_1$ and $\phi$ the optical phase acquired after one round trip in the resonator. We also note $r_1$, $r_2$, $r_{\mathrm{L}}$ and $t_1$, $t_2$, $t_{\mathrm{L}}$ the amplitude reflection and transmission coefficients of respectively $M_1$, $M_2$ and $l$ that we consider without losses. 
The interference between the cavity waves on $M_1$ reads (see Fig.~\ref{cavityGain}):
\begin{figure}[h!]
\includegraphics[width=0.6\linewidth]{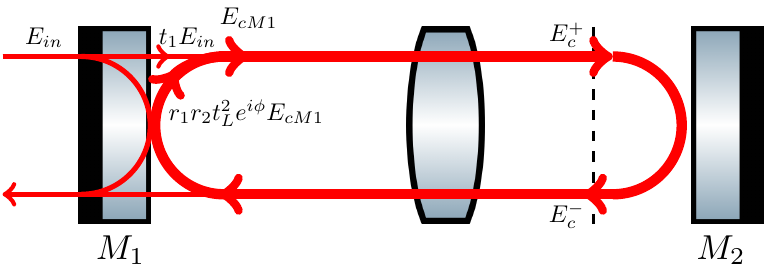}
\centering
\caption{$E_{cM_1}$ is the stationary intra-cavity field propagating in the forward direction on $M_1$. The field $r_1 r_2 t_{\mathrm{L}}^2 e^{i\phi} E_{cM_1}$ is the same field after one round trip inside the cavity. The field $t_1 E_{\mathrm{in}}$ is the incident field leaking inside the resonator.}
\label{cavityGain}
\end{figure}

\begin{equation}\label{CirculatingField}
E_{cM_1}=t_1 E_{\mathrm{in}} + r_1 r_2 t_{\mathrm{L}}^2 e^{i\phi} E_{cM_1}.
\end{equation}
At resonance, $e^{i\phi}$=1, we have therefore:
\begin{equation}
E_{cM_1} =  \dfrac{t_1}{(1 - r_1 r_2 t_{\mathrm{L}}^2)}E_{in}
\end{equation}
which means 
\begin{equation}
P_{cM_1} =  \dfrac{1-r_1^2}{[1 - r_1 r_2 (1-r_{\mathrm{L}}^2)]^2}P_{in}
\end{equation}
Considering that $P_c^-=r_2^2P_c^+$, the amplification factor of the cavity can be written:
\begin{equation}
G=r_2\dfrac{P_c^+}{P_{in}}=r_2\dfrac{P_c^+}{P_{cM_1}}\dfrac{P_{cM_1}}{P_{in}}= r_2(1-r_{\mathrm{L}}^2)\dfrac{P_{cM_1}}{P_{in}}
\end{equation}
We therefore obtain: 

%can be expressed as a function of optics reflection coefficients: %$\mathcal{F} = \frac{FSR}{\delta \nu}$ are

\begin{equation}
G  = \dfrac{r_2(1-r_1^2)(1-r_{\mathrm{L}}^2)}{[1 - r_1 r_2 (1-r_{\mathrm{L}}^2)]^2}. 
\label{eq:Gmax}
\end{equation}
This expression stands for a perfectly aligned resonator. The optical gain will decrease from this maximum value when the cavity is misaligned. We obtain G=100 for the cavity parameters set in Sec. \ref{interro_fields}. 

The Finesse of the resonator can be calculated considering small variations of the frequency around resonance and developing Eq.~\ref{CirculatingField} at first order:
\begin{equation}
\mathcal{F}=\frac{\pi\sqrt{r_1 r_2 (1-r_L^2)}}{1-r_1 r_2 (1-r_L^2)}.
\end{equation}
\section{Details of the numerical calculation of the fields $E_c^\pm(r,z)$}\label{annexB}

We give here details on the convergence of the calculation of the field (Eq.~\eqref{champTOT}).
We iterate the  round trips to a number that equals  four times the finesse $\mathcal{F}$, which allows for a convergence of the field amplitude and phase for all $r$  to less than 1 part in $10^{6}$. More precisely, writing
%\begin{equation}
$E_{[k]}(r,z) = \sum_{n=1}^k E_n(r,z)$
%\end{equation}
and
\begin{align}
\rho_{||}(r,z) &= \frac{\left||E_{[k+1]}(r,z)|-|E_{[k]}(r,z)|\right|}{|E_{[k]}(r,z)|}  \\
\rho_\varphi(r,z) &= \left||\text{arg}(E_{[k+1]}(r,z))| - |\text{arg}(E_{[k]}(r,z))|\right|,
\end{align}
we observe at the end of the calculation that $max_r(\rho_{||}(r,z)) < 10^{-6}$ and $max_r(\rho_\varphi(r,z)) < 10^{-6}$.

In order to lock the cavity, we run the circulating field calculation for a set of different input wavelength and record the gain.
We chose 50 wavelengths evenly spaced in one free spectral range. 
This number allows us to find at least one wavelength on the resonance for finesses up to $\sim 1000$.
We find the wavelength with maximal gain and run the calculation again with a new set of wavelengths around this maximum.
We repeat this process a few times until the wavelength step  $\Delta \lambda$ is below $10^{-9}$ nm.
This number, accounting for the $80$ cm long cavity and $780$ nm wavelength gives us a precision on resonance frequency down to $10^{-5}$ of the free spectral range.
This is  sufficient for Finesse of a few hundreds ($\mathcal{F}=355$ for the cavity parameters set in Sec.~\ref{interro_fields}).

\section{Misalignment of the output mirror}\label{annexC}
\begin{figure}[h!]
\includegraphics[width=1\linewidth]{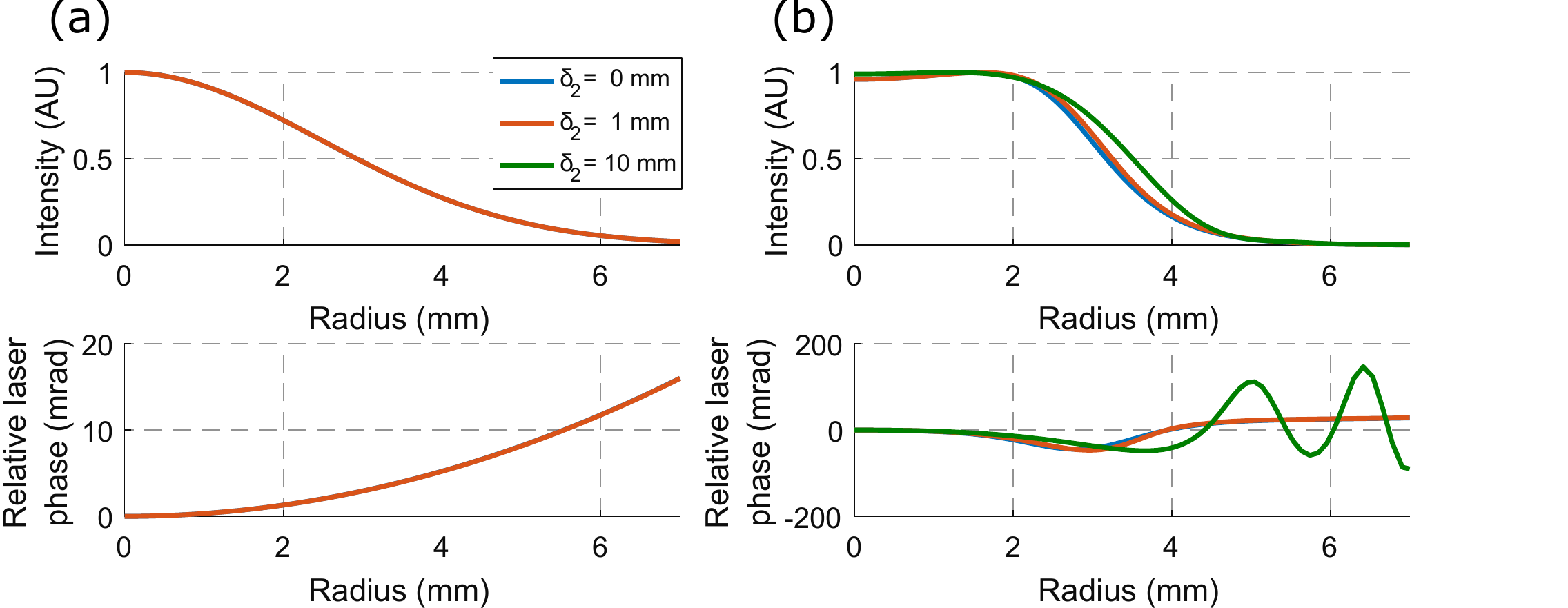}
\centering
\caption{Circulating field in the cavity when misaligning the output mirror $M_2$. The upper plots show the radial intensity profile of the cavity field $E_c^+(r,z)$  at the middle point $z$ between $l$ and $M_2$.
The lower plots shows the radial profile of the relative phase $\Delta\phi(r,z)$ between the two counter-propagating cavity  fields $E_c^+(r,z)$ and $E_c^-(r,z)$. (a) $M_1$ is perfectly aligned, $M_2$ is misaligned by $0$ mm (blue) and $1$ mm (red), the lines are superimposed. (b) $M_1$ is misaligned by $20$ $\mu$m and $M_2$ by $0$ mm (blue), $1$ mm (red) and  $10$ mm (green).}
\label{misalignment_output_mirror}
\end{figure}

Fig.~\ref{misalignment_output_mirror} shows the effect of the misalignment of the output mirror $M_2$ on the interrogation field. When the input mirror $M_1$ is perfectly aligned, the displacement of $M_2$ has nearly no effect on the circulating field (see Fig.~\ref{misalignment_output_mirror}(a)). When $M_1$ is misaligned by a value of $20$ $\mu$m, $M_2$ has to be misaligned by at least a few millimeters to have a significant modification of the circulating field in the cavity. 

\end{document}